\documentclass[a4paper,10pt]{article}

\usepackage{graphicx}%
\usepackage{multirow}%
\usepackage[title]{appendix}%
\usepackage{xcolor}%
\usepackage{textcomp}%
\usepackage{manyfoot}%
\usepackage{booktabs}%
\usepackage{algorithm}%
\usepackage{algorithmicx}%
\usepackage{algpseudocode}%
\usepackage{listings}%

\usepackage[utf8]{inputenc}
\usepackage[T1]{fontenc}
\usepackage[english]{babel}

\usepackage[small,bold]{complexity}
\usepackage{microtype}

\title{Boolean Gates Based on Liquid Marbles}

\author{Luca Cavenaghi, Sandro Erba, Claudio Zandron
\footnote{l.cavenaghi@campus.unimib.it, s.erba9@campus.unimib.it, claudio.zandron@unimib.it}\\
Dipartimento di Informatica, Sistemistica e Comunicazione,\\ Università degli Studi di Milano-Bicocca,\\
Viale Sarca 336, 20126, Milan, Italy}

\date{}

\begin{document}
\maketitle

\begin{abstract}
Liquid Marbles are liquid droplets encapsulated by hydrophobic powder particles.
They offer an efficient approach to handling liquids due to their non-wetting nature.
In this work, starting from the interaction gate proposed in the literature, we describe how
the logic gates AND, XOR, OR, NOT, NAND, and NOR could be realized. Given
the irreversibility and non-conservativeness of classical gates, we also discuss a
possible implementation of the Toffoli gate, a reversible gate, and of the Fredkin
gate, a reversible and conservative gate.
\end{abstract}



\section{Introduction}

Liquid Marbles (LMs) are liquid droplets encapsulated by hydrophobic powders, first introduced in 2001 \cite{bib9}. 

They demonstrate both elastic and liquid properties, and various properties of LMs have been studied in various works. 
For instance, in \cite{Jin2018} authors propose a simple energy relationship to evaluate their coalescence
condition of LMs. 

The coalescence occurs when the kinetic energy surpasses the surface energy of an individual LM, 
leading to particle movement and eventual contact between two colliding LMs at their interface. 
The Weber number serves as a dimensionless value that represents the relative connection between kinetic energy and surface energy:

$$We = \frac{r D v^2}{s}$$

where $r, D, v$, and $s$ are the density, diameter, instantaneous velocity, and surface tension, respectively.

According to the evaluations in \cite{Jin2018}, the kinetic energy of the individual LM should
be at least around $60$ percent of its initial surface energy to make coalescence happen. Otherwise, they bounce elastically and travel following new trajectories.

Properties of Liquid Marbles were also analyzed in \cite{bib10,bib11}, and the optimal composition of their casing (Ni/UHDPE) was proposed so as to obtain marbles that can be controlled by electromagnets due to the presence of Nickel \cite{bib1}. This aspect is crucial in order to define a computing system based on liquid marbles, since the precise timing that controls the movement of the Liquid Marbles is the only way to ensure the proper functioning of such a system. In fact, if timing is not carefully managed, the absence of a Liquid Marble at a proper time, caused by its passing late or early, could be wrongly interpreted as an actual absence of the Liquid Marble.

The computing model based on LMs defined in \cite{bib1} is a collision-based model of computation, that  could be related to the widely discussed Billiard Ball Model (BBM) \cite{bib12}, which deals with hard spheres. However,  LMs behave more similarly in a way described in Soft Sphere Model (SSM), known as the Margolus gate \cite{bib8}. The main difference with this last model lies in the placement of the paths that collect the marbles at the output. In SSM, in fact, it is necessary to give the marbles enough time to collide and bounce elastically before they can travel along a new trajectory. But this is not always the case when dealing with LMs; in fact, depending on their speed, two different types of collision can occur. Experiments performed show that when LMs are traveling at a velocity of 0.21 m/s, then they will bounce off each other, as described in SSM. If, on the other hand, the velocity rises to 0.29 m/s, the two LMs will merge, plummeting with no more angular direction. This can lead to a problem, as the union of two LMs generates a new one with twice the mass, making the LMs in the system no longer identical and indistinguishable from each other. Different masses can then lead to collisions with unpredictable outcomes and other undesirable phenomena. Therefore, it is of paramount importance that the union of two LMs is managed by dividing the new LM produced into two equal parts. This becomes possible with a hydrophobic scalpel, as explained in \cite{bib6}.

Since the velocity of the LMs in the system is, in general, not known, both output channels have been considered, either in the case of elastic rebound or in the case of a merger. It has also been shown that, computationally, the interaction gate can be used as a half-adder; in fact, an implementation of a one-bit full-adder is proposed in \cite{bib1} by arranging two interaction gates in cascade. The Boolean values of the interaction gate are represented by the presence or absence of the LMs at a given point in time. Other circuits have also been proposed with this logic, realizing for example an actuator \cite{bib2}. Interested readers can find further information on the history of liquid-based computations in \cite{bib15}.

In this work, we start by discussing standard logic gates (namely AND, XOR, OR, NOT, NAND and NOR), and we discuss the way to connect them to create logic circuits. 
Then, considering the irreversibility and non-conservativeness of classical gates, we also propose a possible implementation of the Toffoli gate, a reversible gate, and of the Fredkin gate, a reversible and conservative gate. An important consequence of this last result is that, considering the universality of the Fredkin gate, an implementation of universal logic circuits is possible that allow to maintain and reuse the Liquid Marbles initially present in the system: no adding or removal of Liquid Marbles from/to the outside (apart, of course, for those representing the input) are strictly needed to realize computable functions.

The rest of the paper is organized as follows. First, we shortly recall works in the literature related to LMs, both from a biological \cite{bib10,bib11} as well as a computational \cite{bib1} point of view. 
Then, we focus on the implementation of standard logic gates (AND, XOR, OR, NOT, NAND and NOR) using LMs, and we also discuss the implementation of reversible and conservative gates: in particular the Toffoli's gate \cite{bib4} and the Fredkin's gate \cite{bib3}. Finally, we draw some conclusions and propose some further research topics on this subject.

\section{Overview}
\label{sec:Overview}

\begin{figure}[hbt!]
\centering
\includegraphics[width=0.75\textwidth]{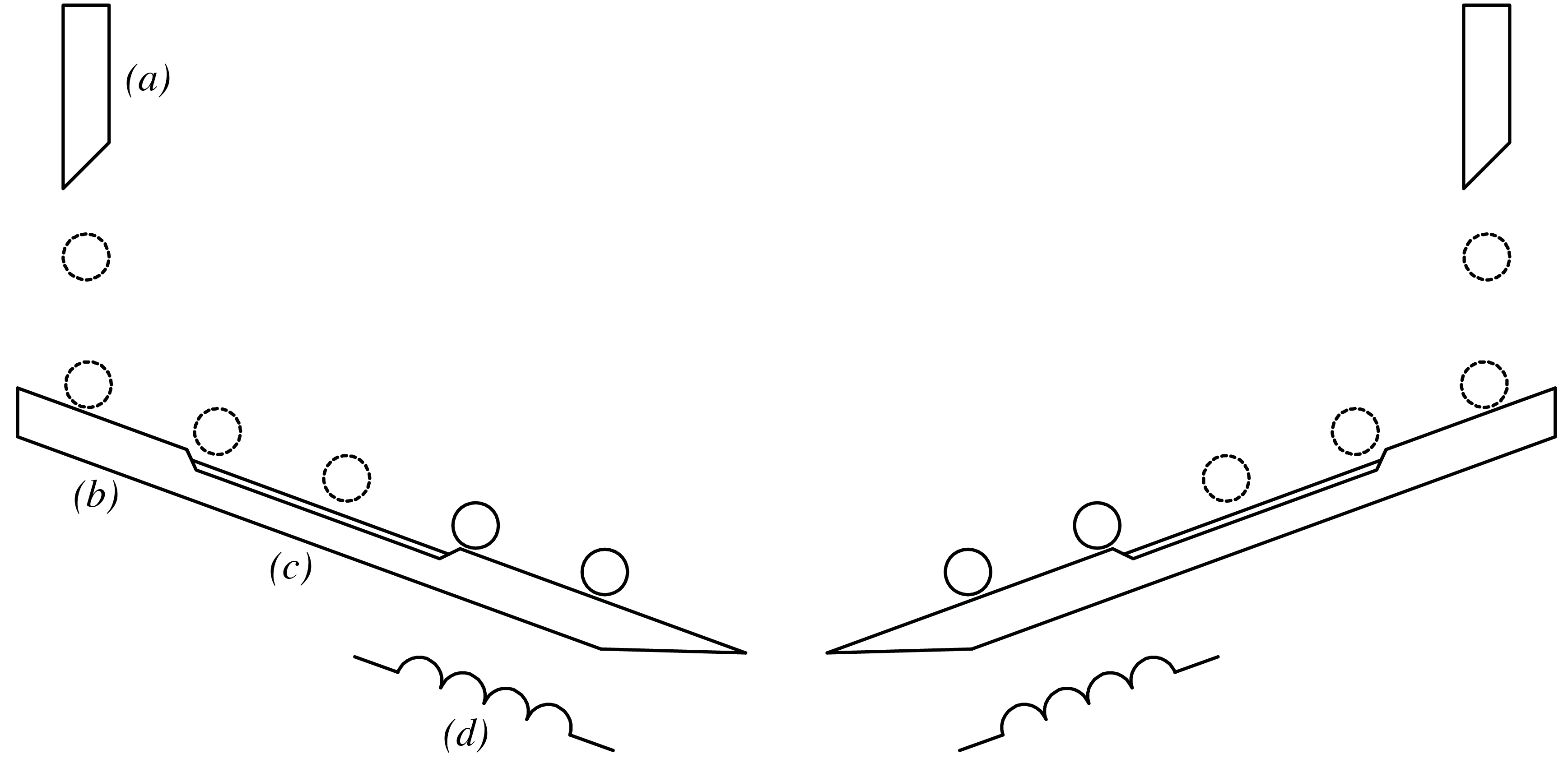}
\caption{\label{fig:IntGate}Interaction gate.}
\end{figure}

The foundation of our work is built upon the interaction gate described in \cite{bib1} and depicted in Fig. \ref{fig:IntGate}; to help the understanding of our work, we recall a brief explanation of the structure of the interaction gate. 

A syringe driver (Fig. \ref{fig:IntGate}(a)) was used to deliver droplets of water at a controlled rate onto a surface treated with a hydrophobic spray (Fig. \ref{fig:IntGate}(b)), causing the water to form beads and roll over a bed of hydrophobic powder (Fig. \ref{fig:IntGate}(c)), resulting in the creation of a continuous stream of LMs with the same properties.

For accurate timing in collision-based computing, an innovative system of electromagnets (EM) was implemented (Fig. \ref{fig:IntGate}(d)). LMs were created using a mix of ultra-high density polyethylene and nickel, providing strength, durability, and a versatile magnetic LM. By positioning an EM behind the acrylic slope, the rolling LMs can be captured and released at will, allowing for control over the timing of collisions. 

The interaction gate was designed to allow for the colliding LMs to have a free path post-collision, enabling the monitoring of LM paths and the creation of a logic gate. Two acrylic pathways were slanted towards each other at 20 degrees and fixed to an acrylic base sheet. The pathways were aligned with a pitch of 38 degrees from the horizontal plane, giving a final slope of 16 degrees from the horizontal plane. Parallel autoformation of hybrid LMs was achieved using the syringe driver, delivering water drops onto the treated section of each slope before rolling across the hydrophobic powder beds to form LMs. The two rolling LMs were then captured using the EMs, allowing for any slight timing deviations to be accounted for. Controlled synchronous or asynchronous release of the EMs results in the LMs simultaneously rolling off the acrylic ramps on collision trajectories.

\section{Description of logic gates}
\label{sec:LogicGates}

\begin{figure}[hbt!]
\centering
\includegraphics[width=1\textwidth]{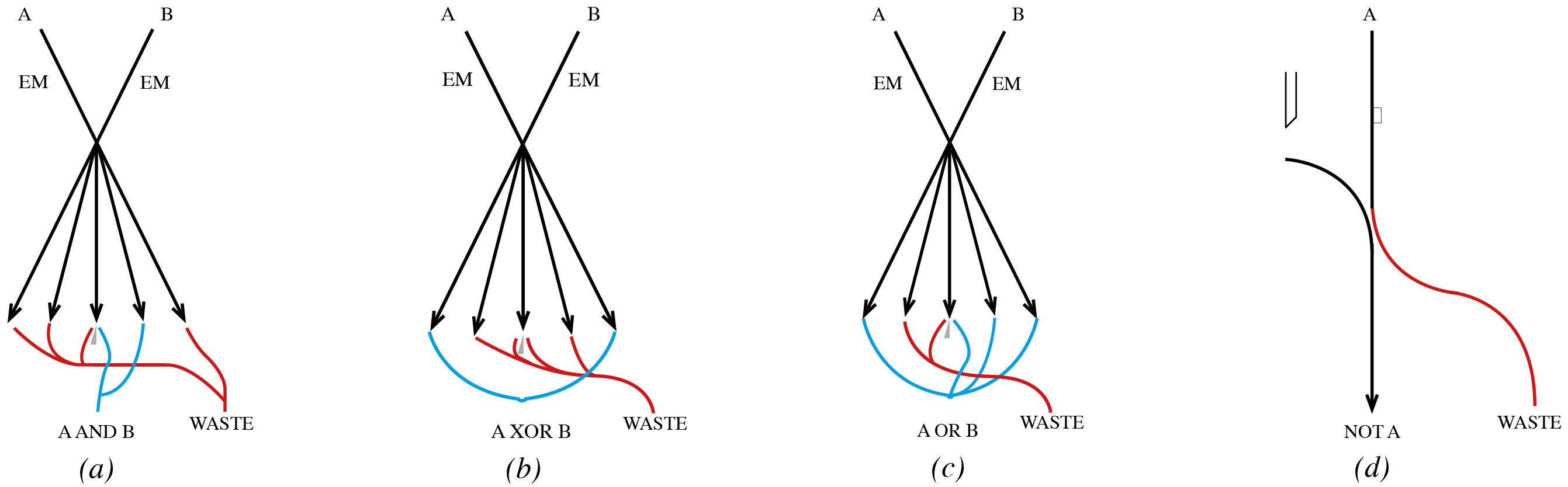}
\caption{\label{fig:GateInizio}AND gate (a), XOR gate (b), OR gate (c), NOT gate (d).}
\end{figure}

In this section, we discuss the realization of a series of fundamental logic gates, which could be used to obtain a computational model based on the collision of Liquid Marbles. All binary gates are based on the interaction gate defined in \cite{bib1} and recalled in the previous section. Channels for passing and addressing LMs were used to design the logic gates. All the fundamental gates that have been developed are depicted in figure \ref{fig:GateInizio}. 

In the following diagrams, the blue-colored paths represent the output of the gates, while the red-colored paths are needed to execute computations, but their output is not used. The gate is synchronized (as explained in the overview) through two electromagnets (EMs), which will allow the LMs to collide or not. Each gate will be discussed in the following, one by one.

\begin{figure}[hbt!]
\centering
\includegraphics[width=1\textwidth]{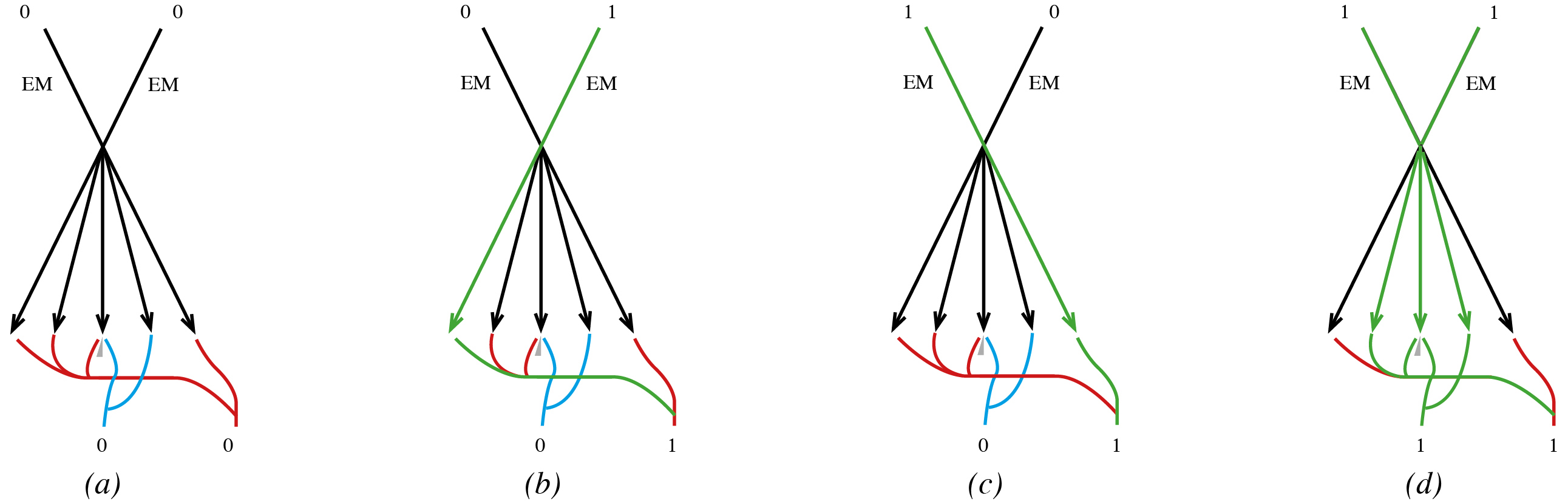}
\caption{\label{fig:AndNuovo}AND gate implementation.}
\end{figure}

In Fig. \ref{fig:AndNuovo} the AND logic gate is depicted. In the AND logic gate, an output LM is obtained only when a collision occurs, that is, when two LMs are present in the input. In order to correctly produce the result, it was necessary to introduce a WASTE channel, where the discarded LMs (that is, LMs present after the interaction of the input LMs that do not correspond to an output of the gate) are addressed.

The figure shows all the possible paths that LMs can take; highlighted in green are the paths that contain an LM, based on each possible input. Those in red are the WASTE channels, where LMs will then be discarded, while in blue are shown the logic gate output paths.
The middle path has both colors as it contains a larger LM that needs to be split by means of a hydrophobic scalpel. This will result in two new LMs being obtained: one will be discarded and brought into the WASTE channel, while the other will go into the output channel. While the two outermost pathways represent the presence of only one input LM, and thus are directed into the WASTE channel, the second and fourth pathways contain LMs that have collided and bounced, not possessing the necessary velocity to merge into a new one. This denotes the presence of both LMs in the input, so one of them will be carried into the output channel, while the other will go into the WASTE channel.
Therefore, to realize this logic AND gate, it suffices to take advantage of the interaction gate discussed earlier, collecting the output LMs as presented with pathways inclined at different angles.

Considering Fig. \ref{fig:AndNuovo}, we have the following. (a) if we have 0-0 as an input (that is, no LMs), then no LM is present as an output. (b) When the input is 0-1, the LM will go through the interaction gate and end up in the leftmost output path; this output path is connected to the WASTE channel which will then have value 1, but the OUTPUT paths of the gate will be empty, thus corresponding to a 0. (c) This case is similar to the previous one. The only difference is that the LM goes to the rightmost path, which is also a WASTE path. (d) When we have an input 1-1, then a collision will occur. We can either have a merge of the two input LMs, thus producing a LM in the central (third from the left) OUTPUT path, or we have a rebound, sending one of the LM to the second path from the left (which is a WASTE path) and the other to the fourth path from the left (which is an OUTPUT path).

It is easy to see that, considering the LMs exiting from the third and the fourth paths, we have the correct output for the AND gate.

\begin{figure}[hbt!]
\centering
\includegraphics[width=1\textwidth]{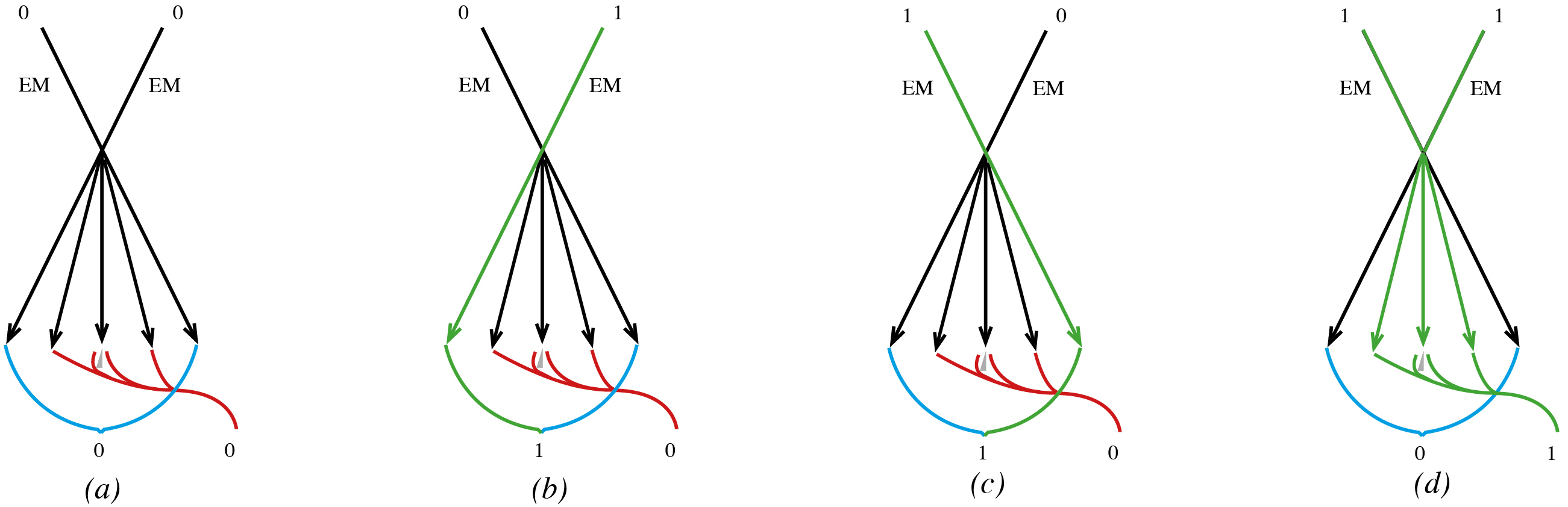}
\caption{\label{fig:XorNuovo}XOR gate implementation.}
\end{figure}

In Fig. \ref{fig:XorNuovo} the XOR logic gate is depicted. In this case, the output channels have been routed in such a way that an output LM is obtained only when exactly a single LM is present in input.
Fig. \ref{fig:XorNuovo} shows all possible paths that the LMs can take (highlighted in green).
When no LMs are given in input, no LM is produced in output. When and input 0-1 is given (or, respectively, 1-0), the LM will cross the interaction gate and end up in the leftmost (respectively, the rightmost) output path, thus correctly producing a 1 as output: see Fig. \ref{fig:XorNuovo} (b) and (c). 
When values 1-1 are given as an input, a collision will occur: the external paths will not contain any LMs, thus obtaining 0 as output (see Fig. \ref{fig:XorNuovo}(d)).

\begin{figure}[hbt!]
\centering
\includegraphics[width=1\textwidth]{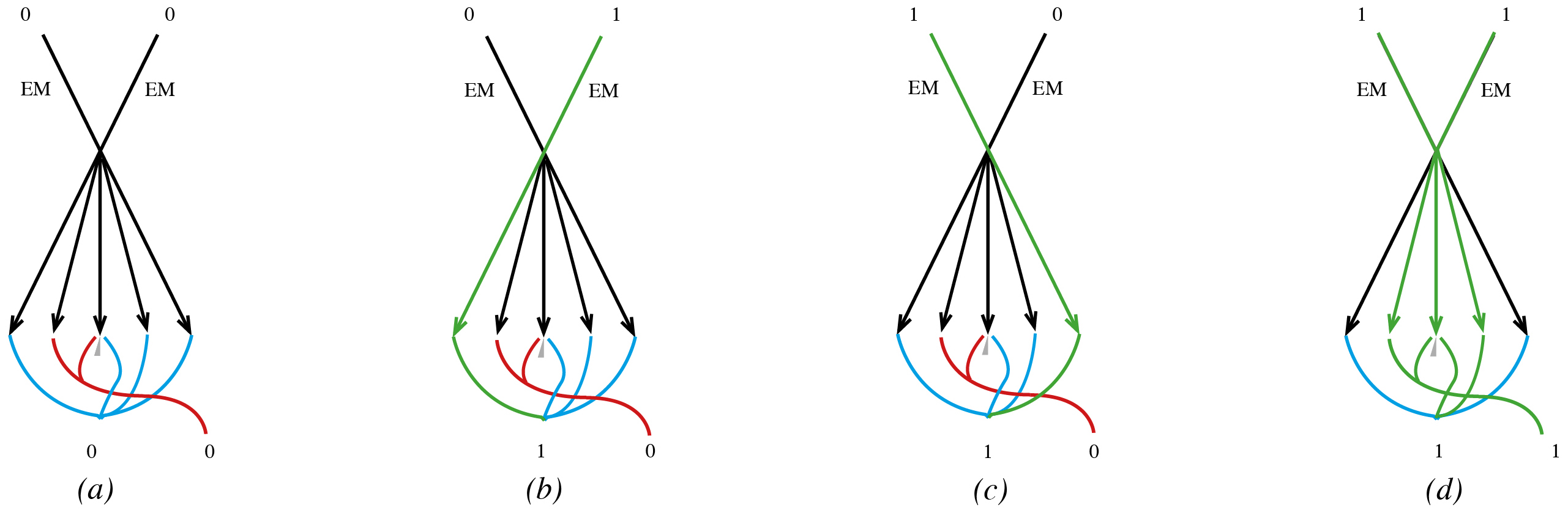}
\caption{\label{fig:OrNuovo}OR gate implementation.}
\end{figure}

In Fig. \ref{fig:OrNuovo} the OR gate is depicted. In the OR logic gate, an LM will be obtained in the output when there is at least one LM in the input. The only case in which no LM will be present in output is when, also in input, no LM is present. In Fig. \ref{fig:OrNuovo} all the possible paths that the LMs can take (highlighted in green) are depicted. When no input is present, no output is produced (Fig. \ref{fig:OrNuovo}(a)). In case of input 0-1 (and similarly, with obvious differences, for the input 1-0) a single LM will cross the interaction gate and it will end up in the leftmost output, thus producing a 1 in output (see Fig. \ref{fig:OrNuovo} (b) and (c)). When an input 1-1 a collision will occur, thus producing an output LM in one of the inner paths (Fig. \ref{fig:OrNuovo}(d)). The gate can thus be obtained by joining all possible output paths.

\begin{figure}[hbt!]
\centering
\includegraphics[width=0.7\textwidth]{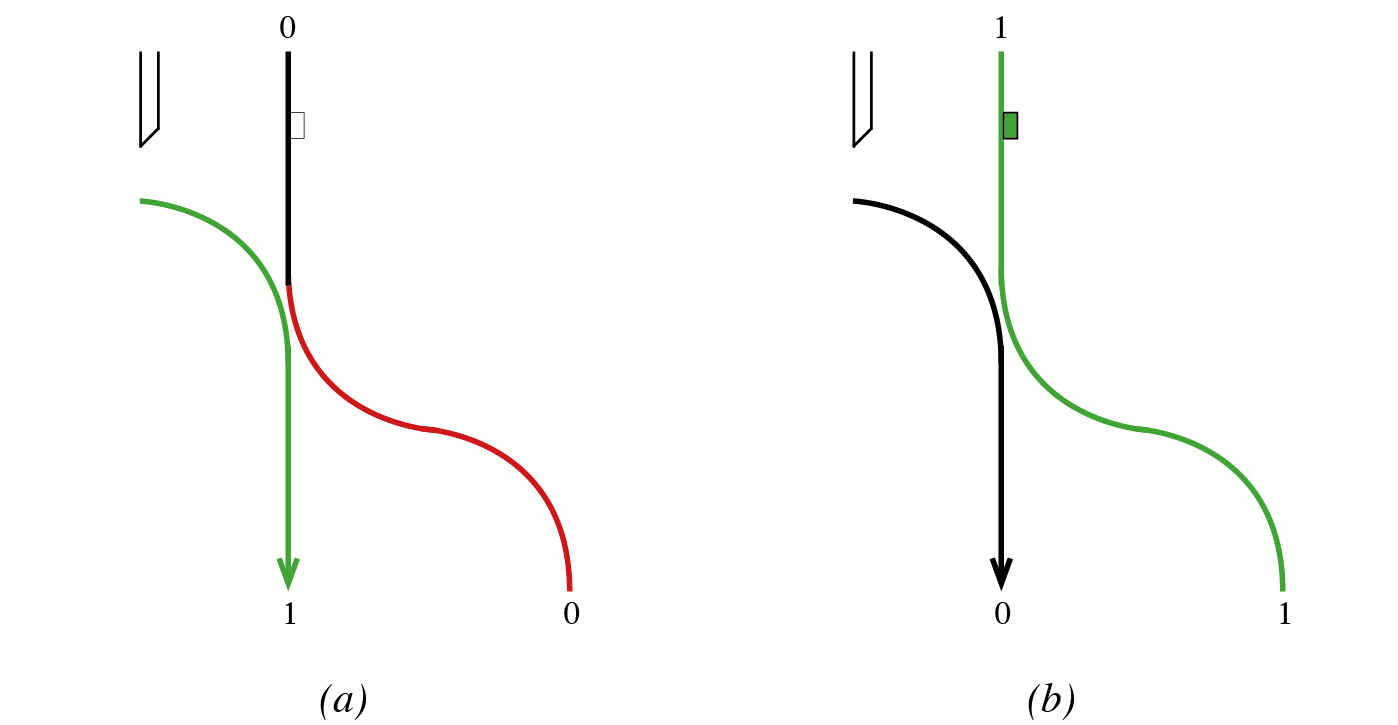}
\caption{\label{fig:NotNuovo}NOT gate implementation.}
\end{figure}

We concentrate now on NOT, NAND and NOR logic gates. We start by describing the implementation of the NOT gate, summarized in Fig. \ref{fig:NotNuovo}. This gate receives a single LM as input, and it reverses the input signal; one possible way to realize it is by adding a syringe \cite{bib1} and a motion sensor. The motion sensor can be seen in Fig. \ref{fig:NotNuovo} and is represented as a small box. If it remains white it means it has not detected any motion, while when it is colored green it means it has detected the passage of an LM. When an LM is given in input (Fig. \ref{fig:NotNuovo}(b)), the motion sensor detects it and the LM is simply routed to the WASTE output. If the gate does not receive an LM in input (Fig. \ref{fig:NotNuovo}(a)), then the motion sensor does not detect movements and a syringe is activated (in a synchronized way with the rest of the circuit) that generates a drop of water that falls on a hydrophobic surface. Going downward, it encounters Ni-UHDPE powder, thus forming an LM in output.

Such a solution requires external hardware to be realized, which could be economically expensive, if several NOT gates are required in a circuit. 

\begin{figure}[hbt!]
\centering
\includegraphics[width=0.8\textwidth]{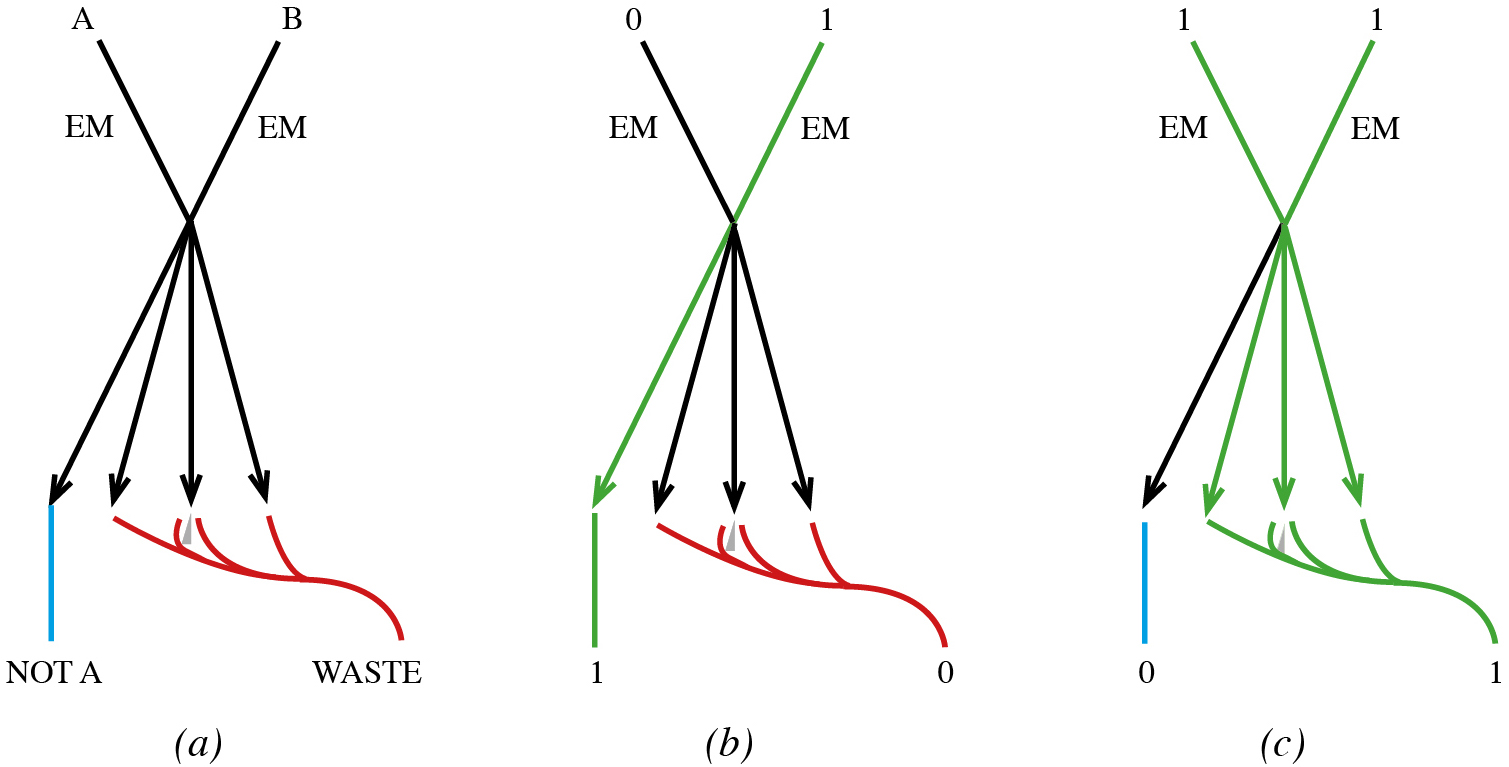}
\caption{\label{fig:NotInteractionExe}Interaction NOT gate implementation.}
\end{figure}

A different solution for the implementation of the NOT gate is thus proposed (see Fig. \ref{fig:NotInteractionExe} (a)) which, like the previous proposed gates, is based on the interaction gate. Such a solution is similar to the one proposed in \cite{bib17} for the Billiard Ball model. The gate has two input paths: the path A is the input value to the NOT gate, while the path B is connected to a synchronized syringe that always produces a control Liquid Marble (B). 

Considering the usual five output paths of the interaction gate, it is easy to see that the rightmost one can be neglected: in fact, since the path B always contains an LM, it is not possible to obtain any LMs in this output path.

Concerning the remaining output path, the leftmost one represent the output of the NOT gate, while the remaining output path are connected to the WASTE.

In the case of an input 0 (Fig. \ref{fig:NotInteractionExe}(b)), no LM will be present in the path A, and we will only have an LM in the input path B. Since no collision will occur, the LM will exit from the leftmost output, thus producing the correct result 1. 

On the contrary, when an LM is present in the input path A, a collision will occur with the LM present in the input path B. No LM will be present in the leftmost output path, thus producing the correct result 0 (Fig. \ref{fig:NotInteractionExe}(c)).

\begin{figure}[hbt!]
\centering
\includegraphics[width=1\textwidth]{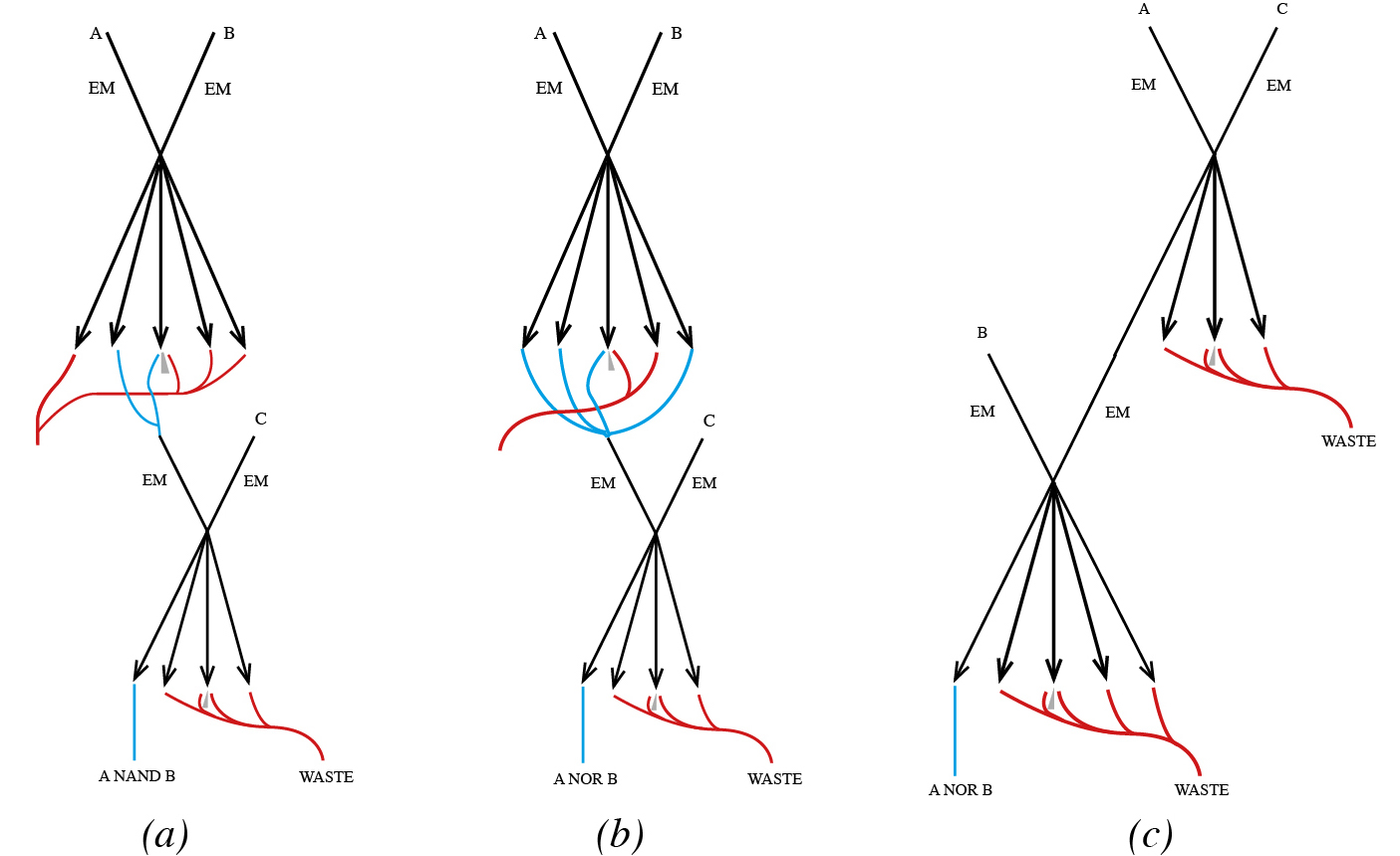}
\caption{\label{fig:NandInteraction}Chained NAND gate (a), chained NOR gate (b), alternative NOR gate (c).}
\end{figure}

\begin{figure}[hbt!]
\centering
\includegraphics[width=1\textwidth]{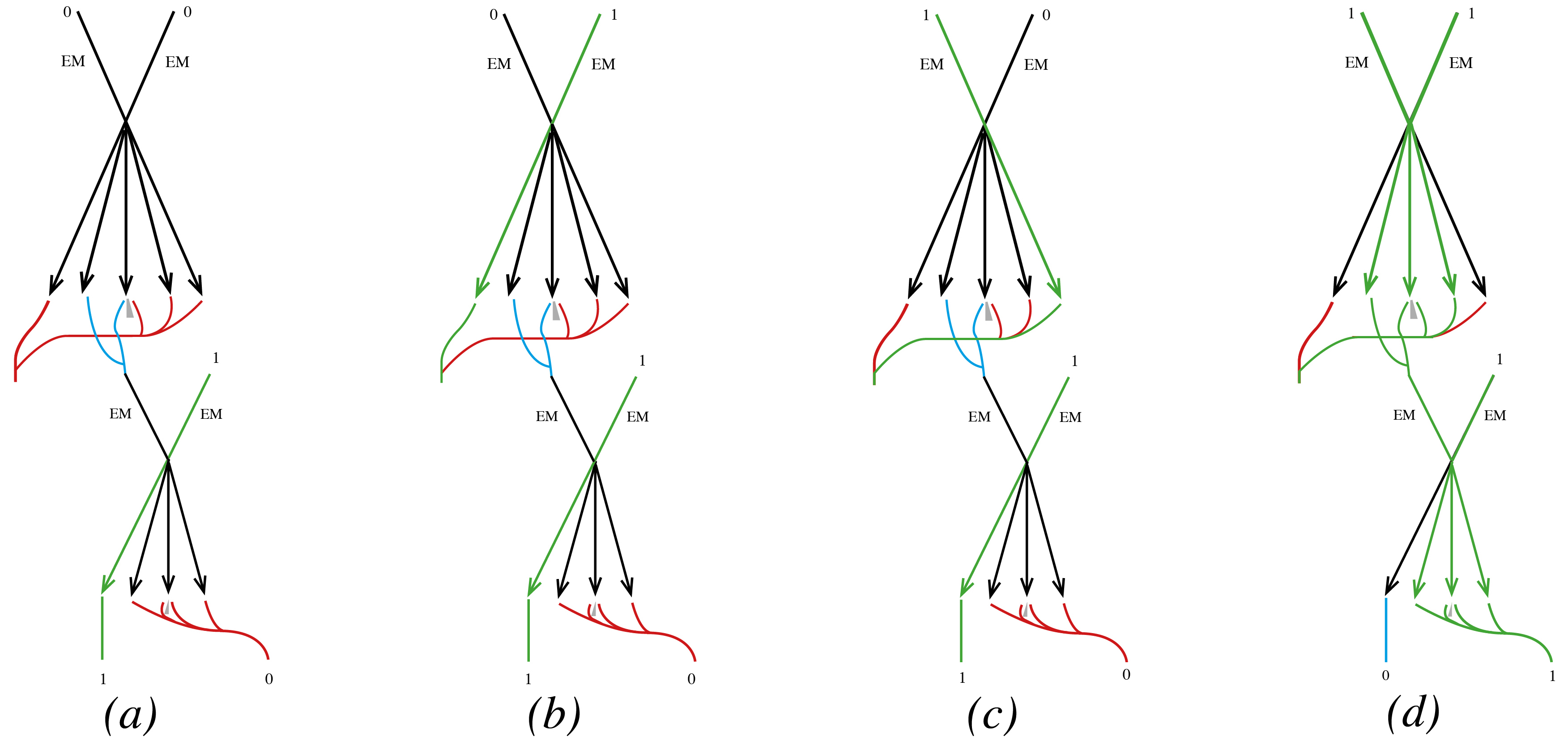}
\caption{\label{fig:NandInteractionExe}Chained NAND gate implementation.}
\end{figure}

In Fig.\ref{fig:NandInteraction}(a) we present the implementation of the NAND gate. To create this gate, we combined the previously described AND gate and NOT gate using an interaction gate. Each approach we tried to implement this gate resulted in a combination of more than one interaction gate: it remains open the problem whether or not it is possible to implement such a gate using a single interaction gate. The description of the computation of the gate is depicted in Fig.\ref{fig:NandInteractionExe}.

\begin{figure}[hbt!]
\centering
\includegraphics[width=1\textwidth]{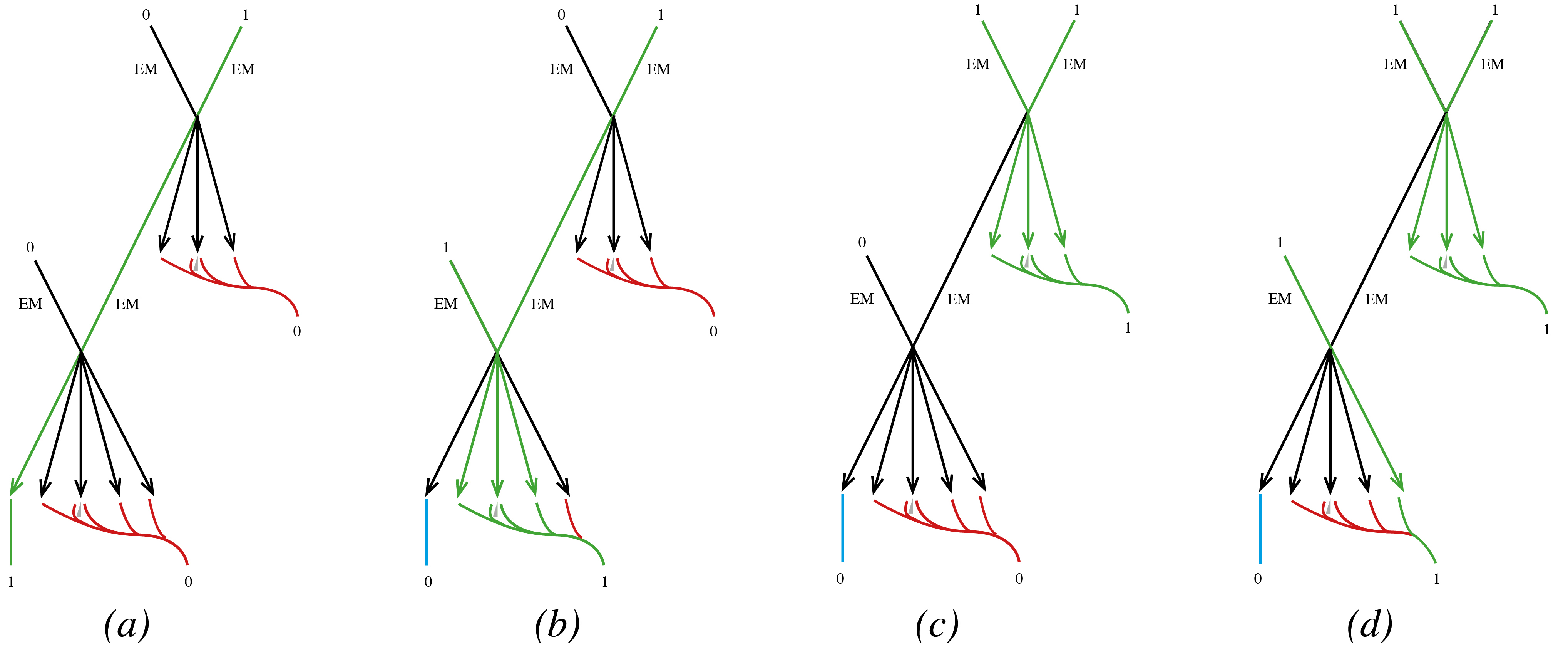}
\caption{\label{fig:NorInteractionExe}Different NOR gate implementation.}
\end{figure}

Fig.\ref{fig:NandInteraction}(b) shows the concatenated NOR gate, which we obtained by combining the previously described OR and NOT gates. Interestingly, we were able to create an alternative version of this gate, as shown in Fig.\ref{fig:NandInteraction}(c). This alternative NOR gate consists of two concatenated NOT gates.
Next, we will describe the computation of the gate in the alternative implementation (Fig.\ref{fig:NorInteractionExe}), since the computation of the first solution is very similar to that of the NAND gate.

In Fig.\ref{fig:NandInteractionExe} all possible paths that the LMs can take (highlighted in green) in an implementation of the NAND gate are depicted. The input C (the control bit of the NOT port) is always equal to 1. When the inputs A and B are 0-0 (Fig.\ref{fig:NandInteractionExe}(a)), 0-1 (Fig.\ref{fig:NandInteractionExe}(b)), and 1-0 (Fig.\ref{fig:NandInteractionExe}(c)), a collision in the AND port will not occur, thus the NOT port will provide 1 in output. On the contrary, when inputs A and B are 1 (Fig.\ref{fig:NandInteractionExe}(d)) a collision will occur, first in the AND port and then in the NOT port, and this will correctly produce a 0 in output.

In Fig.\ref{fig:NorInteractionExe} all the possible paths that the LMs can take (highlighted in green) in an implementation of the NOR gate are depicted. 
Note that the control bit C is located in the upper right corner. The output of the gate is 1 only when both inputs A and B are set to 0 (Fig.\ref{fig:NorInteractionExe}(a)), as C never collides with other LMs in this case, thus producing a 1 as output. Conversely, when the inputs  A and B are 0-1 (Fig.\ref{fig:NorInteractionExe}(b)), 1-0 (Fig.\ref{fig:NorInteractionExe}(c)), or 1-1 (Fig.\ref{fig:NorInteractionExe}(d)), the control bit collides either with A or B, changing its trajectory, and ultimately producing 0 as output.

\section{Reversible and conservative gates}
\label{sec:RevConsGates}

According to Landauer's principle \cite{Landauer1961}, any manipulation of information that is \emph{logically irreversible} leads to an increasing in dissipated heat, and to a reduced ability to extract useful work from the system. 
On the contrary, a \emph{logically reversible} computation could be implemented, at least in principle, without any heat release. A function is said to be \emph{reversible} if its input can always be determined starting from its output, thus when we have a bijective relationship between the input and output states.

This aspect raised a wide interest in the investigation of reversible computing, that was the subject of numerous research works.  
Among them, Fredkin and Toffoli discussed a mathematical model based on conservative logic, proposing two reversible gates: the Toffoli gate and the Fredkin gate \cite{bib3}. 

Considering their importance, we concentrate now on the investigation of a possible implementation of these two gates, using Liquid Marbles and the interaction gate.
We start from the Toffoli gate, a universal and reversible 3-inputs/3-outputs gate, whose truth table is depicted in 
Table \ref{tab:ToffoliGate}.

\begin{table}[hbt!]
\centering
\begin{tabular}{|c|c|c|c|c|c|}
\hline
c & x\textsubscript{1} & x\textsubscript{2} & y & g\textsubscript{1} & g\textsubscript{2} \\
\hline
0 & 0 & 0 & 0 & 0 & 0 \\
\hline
0 & 0 & 1 & 0 & 0 & 1 \\
\hline
0 & 1 & 0 & 0 & 1 & 0 \\
\hline
0 & 1 & 1 & 0 & 1 & 1 \\
\hline
1 & 0 & 0 & 1 & 0 & 0 \\
\hline
1 & 0 & 1 & 1 & 0 & 1 \\
\hline
1 & 1 & 0 & 1 & 1 & 1 \\
\hline
1 & 1 & 1 & 1 & 1 & 0 \\
\hline
\end{tabular}
\caption{Toffoli gate truth table.}
\label{tab:ToffoliGate}
\end{table}

\begin{figure}[hbt!]
\centering
\includegraphics[width=0.5\textwidth]{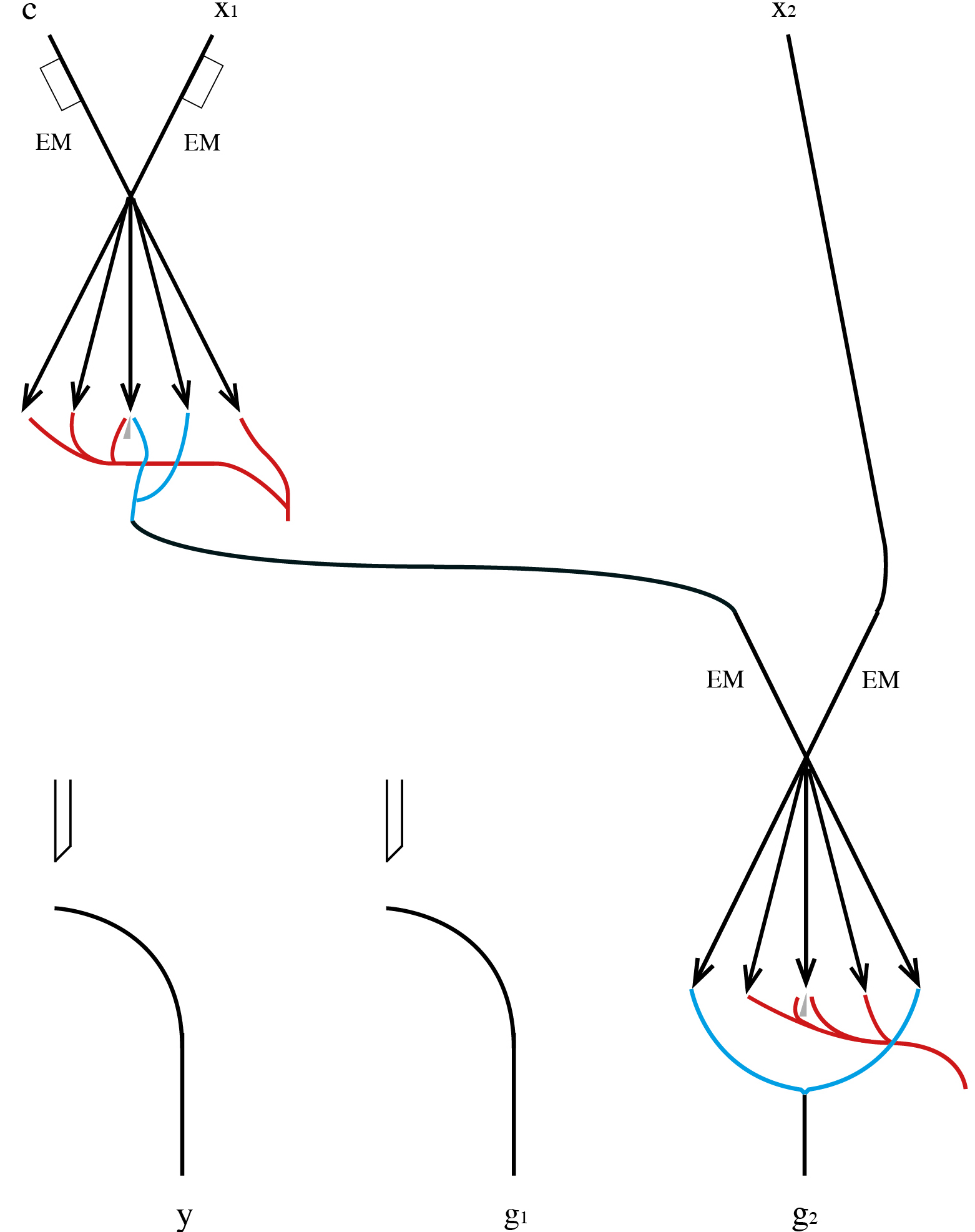}
\caption{\label{fig:ToffoliUniti1}An implementation of the Toffoli gate.}
\end{figure}

A first possible implementation of the Toffoli gate is depicted in Fig. \ref{fig:ToffoliUniti1}, and it is obtained by means of a composition of an AND gate and a XOR gate, in succession. The gate produces three outputs: y, g\textsubscript{1} and g\textsubscript{2}. g\textsubscript{2} is obtained by performing an XOR operation between the result of a logical AND operation between inputs c and x\textsubscript{1}, and input x\textsubscript{2}. In addition to these, the gate also outputs copies of inputs c and x\textsubscript{1}, which are labeled as y and g\textsubscript{1}, respectively.

\begin{figure}[hbt!]
\centering
\includegraphics[width=0.5\textwidth]{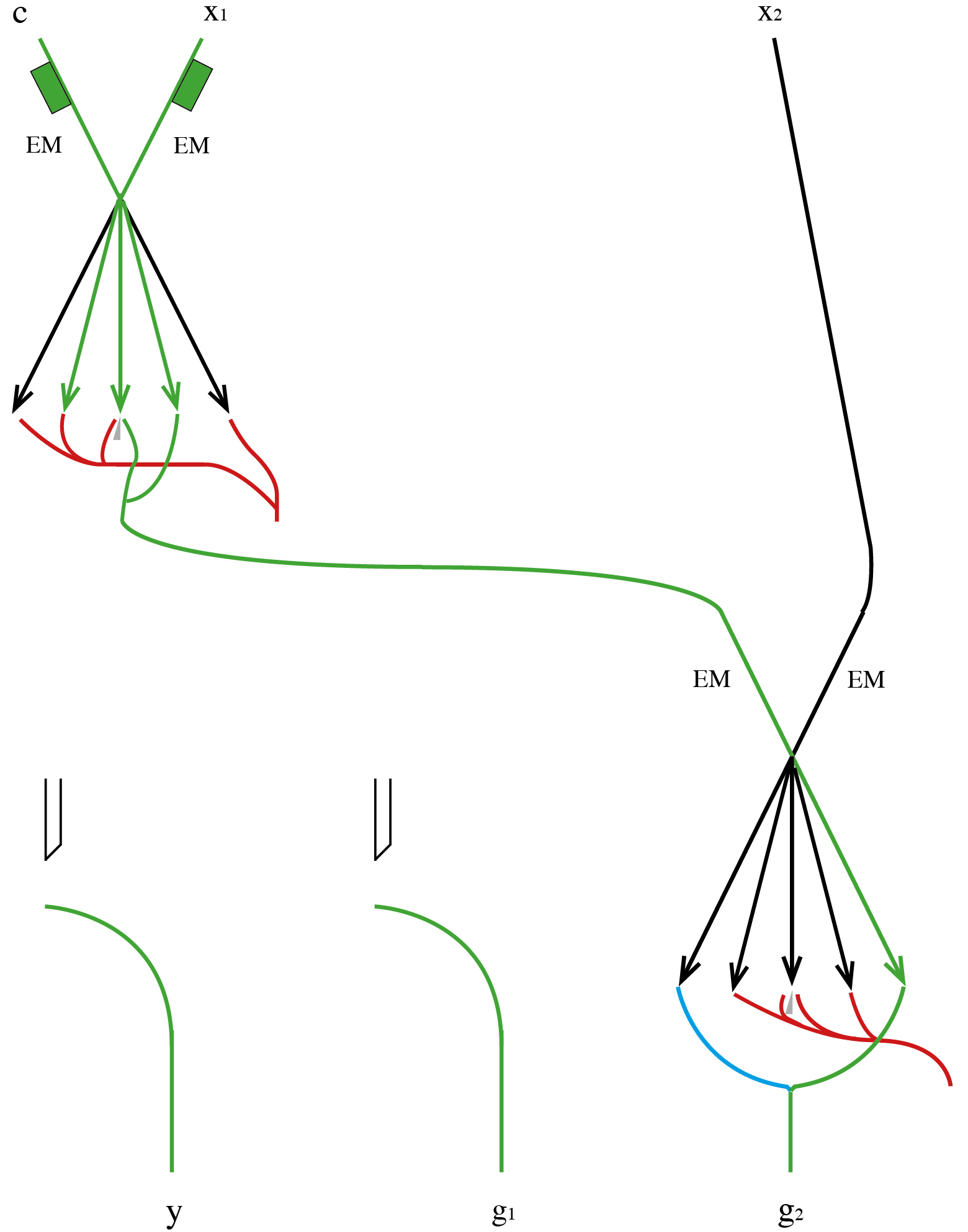}
\caption{\label{fig:ToffoliUniti2} An example of the functioning of the Toffoli gate.}
\end{figure}

An example of functioning of a simulation of the Toffoli gate for the input c=1, x\textsubscript{1}=1, x\textsubscript{2}=0 is illustrated in Fig. \ref{fig:ToffoliUniti2}, in order to clarify its functioning.
With input c=1, x\textsubscript{1}=1, x\textsubscript{2}=0, the LMs enter the first AND gate that generates output 1, then pass through the underlying XOR gate that generates output 1. Outputs y and g\textsubscript{1} are copied from their respective inputs. The operations for the other possible inputs are similar.

It is clear that such a solution has the problems already highlighted in the previous section for other logic gates: we need to add further elements to the interaction gate such as injection mechanisms and movement sensors and, moreover, we need to add new LMs from outside the gate when needed (thus, with the need to arrange an external storage), or removing them from the circuit when appropriate (thus wasting them or collecting them in an external storage).

One possible way to avoid this, is by an implementation of the Fredkin gate, which is not only universal and reversible, but it is also \emph{conservative}. Conservative computing models mirrors more closely physics with respect to traditional models, by an efficient use of the "computing resources" actually offered by nature. In particular, conservative logic shows that it is ideally possible to build logic circuits with zero internal power dissipation. 
Using conservative logic, it is possible to preserve the computational capabilities of ordinary digital logic while satisfying the "physical" constraints of reversibility and conservation \cite{bib3}.

A notable feature of the Fredkin gate, is that the number of bits set to 1 in input and output is always the same, as one can see from its truth table in Table \ref{tab:FredkinGate}. As a consequence, in circuits developed using LMs this corresponds to maintaining the amount of LMs inside the gate: the number of LMs in output will always be equal to the number of LMs in input. This computation element can be viewed as a device that conditionally crossovers two data signals based on the value of a control signal. When this value is 1, the two data signals follow parallel paths; when the value is 0, they cross.

\begin{table}[hbt!]
\centering
\begin{tabular}{|c|c|c|c|c|c|}
\hline
u & x\textsubscript{1} & x\textsubscript{2} & v & y\textsubscript{1} & y\textsubscript{2} \\
\hline
0 & 0 & 0 & 0 & 0 & 0 \\
\hline
0 & 0 & 1 & 0 & 1 & 0 \\
\hline
0 & 1 & 0 & 0 & 0 & 1 \\
\hline
0 & 1 & 1 & 0 & 1 & 1 \\
\hline
1 & 0 & 0 & 1 & 0 & 0 \\
\hline
1 & 0 & 1 & 1 & 0 & 1 \\
\hline
1 & 1 & 0 & 1 & 1 & 0 \\
\hline
1 & 1 & 1 & 1 & 1 & 1 \\
\hline
\end{tabular}
\caption{Fredkin gate truth table.}
\label{tab:FredkinGate}
\end{table}

Since the Fredkin gate is a universal gate, any function can be calculated by means of a logic circuit using only such gates. This means that if it is possible to implement such gate using LMs and interactions gates only, then we could completely avoid the need of adding/collect LMs from/to external storages. As a consequence, this would result in a simplified circuit, not requiring external elements like injectors, movement sensors, or storages.

\begin{figure}[hbt!]
\centering
\includegraphics[width=0.8\textwidth]{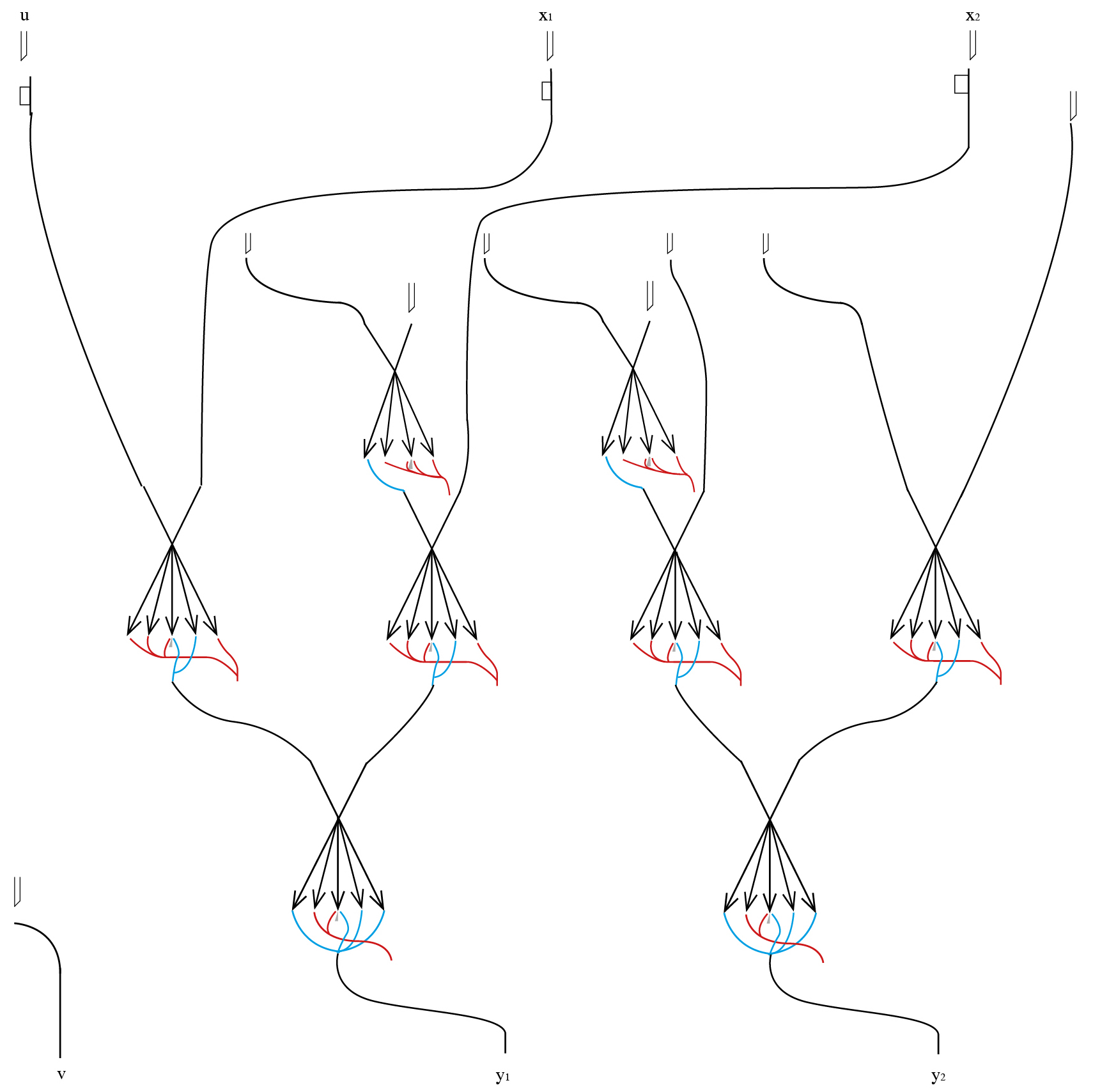}
\caption{\label{fig:FredkinGate} A first implementation of the Fredkin gate.}
\end{figure}

In the following we will show that, in fact, this is the case. We start by presenting a Fredkin gate implemented using "standard" gates, developed in the previous section, which is depicted in Fig. \ref{fig:FredkinGate}.
As one may notice, the gate still contain many syringes and motion sensors; furthermore, the developed Fredkin gate does not maintain conservativity internally, because it is formed by gates that do not satisfy this property.

\begin{figure}[hbt!]
\centering
\includegraphics[width=0.5\textwidth]{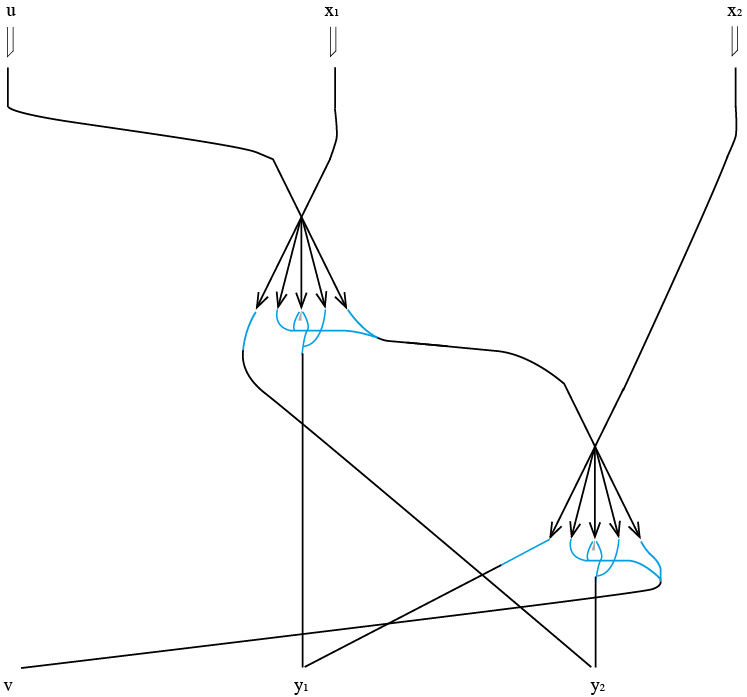}
\caption{\label{fig:FredkinUniti1} A direct Fredkin gate implementation.}
\end{figure}

To avoid the problems of such a solution, a "direct" implementation is needed, that allow to avoid the use of external instruments and LMs. One possible such implementation, that can be obtained by using two interaction gates, is depicted in Fig.\ref{fig:FredkinUniti1}. As one can see, the obtained gate do not need motion sensors and syringes, and the number of LMs in input is the same as the number of LMs in output. 

\begin{figure}[hbt!]
\centering
\includegraphics[width=1\textwidth]{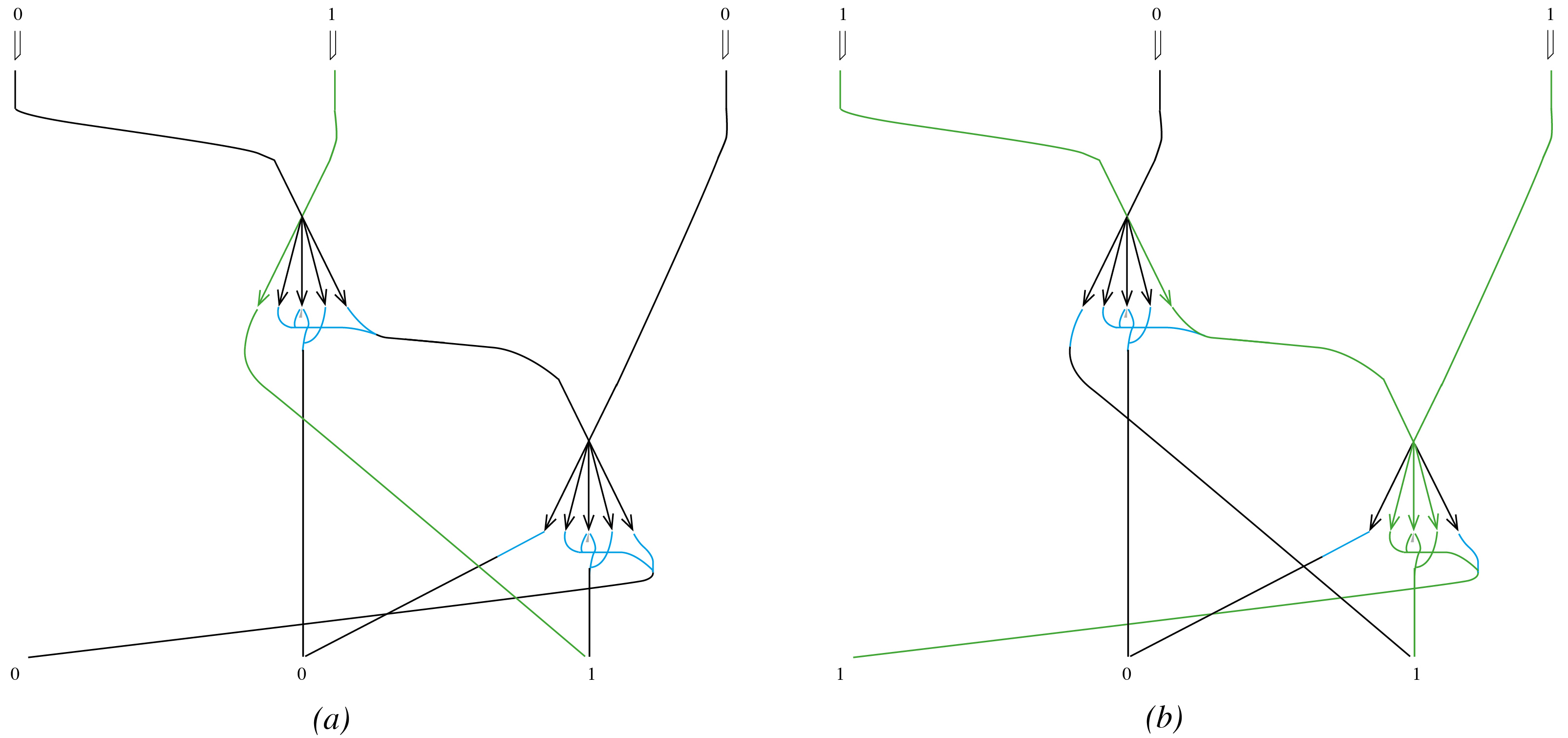}
\caption{\label{fig:FredkinUniti2}Direct Fredkin gate 0-1-0 (a) and 1-0-1 (b) simulation.}
\end{figure}

Fig.\ref{fig:FredkinUniti2}(a) shows an example of how this gate operates. In this example, the input values are u=0, x\textsubscript{1}=1, and x\textsubscript{2}=0. Since u is 0 (the control bit), the Fredkin gate will swap the values of $x1$ and $x2$ during execution. The LM starts at $x1$ and passes through the first gate without collision occurring; this gate diverts the trajectory to the output $y2$. As a result, the output values will be v=0, y\textsubscript{1}=0, and y\textsubscript{2}=1.

We show in Fig.\ref{fig:FredkinUniti2}(b) another example in which LM collision occurs inside the gate. In this example, the input values are u=1, x\textsubscript{1}=0, and x\textsubscript{2}=1. Since u is 1 (the control bit), Fredkin's gate will not swap the values of $x1$ and $x2$ during execution.  In this case there will be a LM collision in the rightmost gate, generating the output values v=1, y\textsubscript{1}=0, and y\textsubscript{2}=1. The other possible input combinations are trivial, so we will not show them in detail.

As stated before, this result allow to prove the possibility to build logic circuits without the need of added instruments (apart from those required to implement and synchronise the interactions gates), and with an amount of LMs that can be defined at the beginning of computation. There is no need to add or remove LMs from the circuits during the computations:  some added LMs could only be required in some cases to replace degraded LMs.

\section{Conclusions}
\label{sec:Conclusions}

We proposed a possible implementation of various boolean gates based on the collision of Liquid Marbles. 
Building on the interaction gate proposed in \cite{bib1}, a set of "standard" logic gates (AND, XOR, OR, NOT, NAND, and NOR) were implemented. 

We also considered reversible gates, in particular the Toffoli gate and the Fredkin gate. 
In particular, we showed that an implementation of the Fredkin gate can be obtained by simply combining two interaction gates. Considering the universality of the Fredkin gate, this proved  the possibility to build logic circuits based on LMs that do not need added instruments (apart from those required to implement and synchronise the interactions gates), and using an amount of Liquid Marbles that can be defined at the beginning of computation, without the need of adding or removing them from the circuit during the computations.

Dealing with Liquid Marbles, we have some advantages not found in traditional circuits. For instance, we could consider the use of marbles of different sizes, and with a careful control, it could be possible to associate different diameters with different values, breaking out of the limitations of a binary system.
Moreover, since they are in fact composed of a liquid, computation can be viewed as the transport of a load in different parts of the circuit, which can be exploited in disparate ways, such as initiating chemical reactions within the marble due to their coalescence \cite{bib14}.

This research activity aimed to provide a solid basis for implementing in the future a series of logic circuits, that will form a complete datapath.

\section*{Acknowledgements}
This work has been partially supported by Universit\`a degli Studi di Milano-Bicocca, Fondo di Ateneo Quota Dipartimentale (2019-ATE-0454).


\begin{thebibliography}{99}

\bibitem{bib12}
Adamatzky, A., and J. Durand-Lose. "Collision-based computing." Handbook of natural computing. Springer, Berlin, Heidelberg, 2012. 1949-1978.

\bibitem{bib5}
Adamatzky, A. "Binary full adder, made of fusion gates, in a subexcitable Belousov-Zhabotinsky system." Physical Review E 92.3 (2015): 032811.

\bibitem{bib15}
Adamatzky, A. "The dry history of liquid computers." arXiv preprint arXiv:1811.09989 (2018).

\bibitem{bib9}
Aussillous, P., and Qu\' er\' e, D. "Liquid marbles." Nature 411.6840 (2001): 924-927.

\bibitem{bib11}
Aussillous, P., and Qu\' er\' e, D. "Properties of liquid marbles." Proceedings of the Royal Society A: Mathematical, Physical and Engineering Sciences 462.2067 (2006): 973-999.

\bibitem{bib6}
Bormashenko, E., and Bormashenko, Y. "Non-stick droplet surgery with a superhydrophobic scalpel." Langmuir 27.7 (2011): 3266-3270.

\bibitem{bib1}
Draper, T.C., Fullarton, C., Phillips, N., de Lacy Costello, B.P.J., Adamatzky, A., "Liquid marble interaction gate for collision-based computing." Materials Today 20.10 (2017): 561-568.

\bibitem{bib2}
Draper, T.C., Fullarton, C., Phillips, N., de Lacy Costello, B.P.J., Adamatzky, A., "Liquid marble actuator for microfluidic logic systems." Scientific reports 8.1 (2018): 1-9.

\bibitem{bib3}
Fredkin, E., and Toffoli, T. "Conservative logic." International Journal of theoretical physics 21.3 (1982): 219-253.

\bibitem{bib7}
Fullarton, C., Draper, T.C., Phillips, N., Mayne, R., de Lacy Costello, B.P.J., Adamatzky, A., "Evaporation, lifetime, and robustness studies of liquid marbles for collision-based computing" Langmuir 34.7 (2018): 2573-2580.

\bibitem{bib17}
Hosseini, Hadi, and Gerhard W. Dueck. "Toffoli gate implementation using the billiard ball model." 2010 40th IEEE International Symposium on Multiple-Valued Logic. IEEE, 2010.

\bibitem{Jin2018} Jin, J., Ooi, C.H., Dao, D.V., Nguyen, N.T., Liquid marble coalescence via vertical collision, Soft Matter, 14, 2018, 14, 4160-4168.

\bibitem{bib10}
Khaw, M.K., Ooi, C.H., Mohd-Yasin, F., Vadivelu, R., John, J., Nguyen, N.T., "Digital microfluidics with a magnetically actuated floating liquid marble." Lab on a Chip 16.12 (2016): 2211-2218.

\bibitem{Landauer1961} Landauer, R. "Irreversibility and Heat Generation in the
Computing Process", \emph{IBM Journal of Research and Development}, 5, 1961, 183--191.

\bibitem{bib8}
Margolus, N. "Universal cellular automata based on the collisions of soft spheres." Collision-based computing (2002): 107-134.

\bibitem{bib4}
Toffoli, T. "Reversible computing." International colloquium on automata, languages, and programming. Springer, Berlin, Heidelberg, 1980.

\bibitem{bib14}
Chen, Z., Zang, D., Zhao, L., Qu, M., Li, X., Li, L., Geng X.,
Liquid Marble Coalescence and Triggered Microreaction Driven
by Acoustic Levitation, Langmuir, 33(25), 2017, 6232--6239.


\end{thebibliography}
\end{document}